\documentclass[%
reprint,
 amsmath,amssymb,
 aps,showkeys,
pre
]{revtex4-1}

\usepackage{graphics}
\usepackage{graphicx}% Include figure files
\usepackage{dcolumn}% Align table columns on decimal point
\usepackage{bm}% bold math
\usepackage{hyperref}% add hypertext capabilities
\usepackage{placeins}

\begin{document}

\title{Inelastic collapse and near-wall localization of randomly accelerated particles }
\date\today

\author{S. Belan, A. Chernykh, V. Lebedev, G. Falkovich}
%\email{sergb27@yandex.ru}
\affiliation{Moscow Institute of Physics and Technology, 141700 Dolgoprudny, Russia}
\affiliation{Landau Institute for Theoretical Physics,
 142432 Chernogolovka, Russia}
\affiliation{Institute of Automation and Electrometry,  Novosibirsk, Russia}
\affiliation{Novosibirsk State University,Novosibirsk, Russia}
\affiliation{Weizmann Institute of Science, Rehovot,  Israel}
\affiliation{Institute for Information Transmission Problems, Moscow, Russia}

 \begin{abstract}
 The inelastic collapse of stochastic trajectories of a randomly accelerated particle moving in half-space $z > 0$ has been discovered by McKean and then independently
re-discovered by Cornell et. al. 
The essence of this
phenomenon is that particle arrives to a wall at $z = 0$ with zero velocity after an infinite number of inelastic collisions if the restitution coefficient $\beta$ of particle velocity is smaller than the critical value $\beta_c=\exp(-\pi/\sqrt{3})$.
%The essence of this
%phenomenon is that there is a finite probability that in a finite time an infinite number of inelastic collisions makes  particle to settle with zero velocity on the wall if the velocity restitution coefficient $\beta$ is smaller than the critical value $\beta_c=\exp(-\pi/\sqrt{3})$.
We demonstrate that inelastic collapse %[Cornell et al. Phys. Rev. Lett. 81, 1142 (1998)]
takes place also in a wide class of models with spatially inhomogeneous random force and, what is more, that the critical value $\beta_c$ is universal. 
That class includes an important case of inertial particles in wall-bounded random flows. 
To establish how the inelastic collapse influence the particle distribution, we construct an exact equilibrium probability density function  $\rho(z,v)$ for particle position and velocity.
The equilibrium distribution exists only at $\beta<\beta_c$ and indicates that inelastic collapse does not necessarily mean the near-wall localization.
 \end{abstract}

% \ocis{240.6680, 160.3918, 260.3910}

\maketitle

%\section{Introduction}
%\label{sec:INTRODUCTION }

\section{INTRODUCTION}

The models of a randomly accelerated particles which collide inelastically with a wall arise naturally in different contexts ranging from driven granular matter to dynamics of confined polymers (see e.g. \cite{Burkhardt_2014} and relevant references therein).
In the simplest one-dimensional model, % without viscous damping,
 a particle is subjected to Gaussian white noise and instantaneously loses a certain part of its velocity at the moments of reflection from the wall.
The first rigorous results for this problem go back to McKean \cite{McKean}, who discovered that trajectory tends to touch the origin of phase space (particle velocity turns zero on the wall) if the restitution coefficient of particle velocity is smaller that some critical value.
Subsequently, this phenomenon was named the inelastic collapse and has been extensively studied in physics \cite{Cornell, Swift_1999, Florencio_2000, Burkhardt_2000_1, Burkhardt_2000_2,  Anton_2002, Burkhardt_2004} and mathematics \cite{Jacob_2012, Hwang_2015} literature.
In particular, it has been established that  
%a particle that finds itself on the wall with zero velocity does not necessarily remain on the wall forever. 
inelastic collapse does not necessarily force the particle to remain at the wall forever \cite{ Anton_2002, Burkhardt_2004,Jacob_2012, Hwang_2015}.

More recently there has been significant interest in specific generalization of the aforementioned classical model, which includes viscous damping together with inhomogeneous noise whose intensity increases with distance to the wall.
First of all, the inhomogeneous generalization is relevant to the single-particle dynamics in turbulent environment.
Say, for an inertial particle in wall-bounded turbulence, the intensity of noise term, which is proportional to
the fluctuating velocity of the carrier fluid, vanishes at the wall due to no-slip
boundary condition.
Another fluid-dynamical application of the model with inhomogeneous random force describes the relative motion of inertial particles in turbulent flows.
Surprisingly, it was found that the critical restitution coefficient, corresponding to the inelastic collapse in the classical homogeneous model, controls also the localization-delocalization transition for the inertial particle in viscous sublayer of wall-bounded turbulence \cite{Belan_2015} and path-coalescence transition in relative motion of particles with  very large inertia \cite{Belan_2016}.

These findings have motivated us to consider here the behaviour of a particle which is accelerated by a random force with arbitrary intensity profile.
We demonstrate that some results, which was previously derived within homogeneous model, are valid also for more general situation.
In particular, phenomenon of inelastic collapse turns out to be universal, although
the time needed for collapse can be either finite or infinite depending on the specific form of inhomogeneity.
Moreover, we construct an exact equilibrium probability distribution for particle velocity and coordinate which is a zero-flux solution of the Fokker-Planck equation with inhomogeneous temperature.
That distribution corresponds to the non-trapping boundary condition at the origin of phase space, i.e. the particle that finds itself on the wall with zero velocity does not stay there for later times. 
The main focus of the present work is on the localization properties of the particle in the non-trapping case: does it remain in vicinity of the wall or escape to infinity?
As explained below, the answer depends on the elasticity of particle-wall collision
and on the spatial profile of noise intensity.
The analysis includes the classical problem with a homogeneous noise as a particular case.
At the end, we discuss how the presence of viscous damping affect the main conclusions of our analysis.

\section{MODEL FORMULATION}

Consider the motion of a particle in one-dimensional domain
$z\ge0$ under an influence of random force with spatially inhomogeneous intensity.
The particle coordinate and velocity evolve in time accordingly to the following equations
\begin{equation}
\label{equation of motion}
\frac{dz}{dt}=v,\ \ \ \ \frac{dv}{dt}=\xi,
\end{equation}
in which the Gaussian noise $\xi$ is delta-correlated in time with a zero mean and the pair correlation function
\begin{equation}
\langle\xi(z,t_1)\xi(z,t_2)\rangle=2\kappa(z)\delta(t_1-t_2).
\end{equation}
The noise magnitude $\kappa(z)$ can be interpreted as a non-uniform temperature or turbulence intensity.
While many of the results presented here are valid for quite arbitrary $z$-dependence of $\kappa$, we mainly focus our attention on power-law model
%Here we concentrate mainly on the case of power-law temperature distribution
\begin{equation}
\label{temperature_profile}
\kappa(z)= z^m, \ \ \ \ m\ge 0,
\end{equation}
where index $m$ can be called the inhomogeneity degree.
The uniform case $m=0$ has been intensively discussed in the context of driven granular matter \cite{Burkhardt_2014}.
The model with $m=2$ describes the motion of a single particle near the minimum of turbulence intensity \cite{Belan_2014} and the relative motion of two close particles in a one-dimensional random flow \cite{Gustavsson_2011,Belan_2016}, both in the limit of high inertia.
% serves as a generic profile of the temperature near a minima of noise intensity.
At $m=4$ and $m=1$ the profile (\ref{temperature_profile}) mimics the space-dependent turbulent diffusivity of wall-bounded turbulent flow in the viscous sub-layer \cite{Belan_2015} and in the  logarithmic layer \cite{MY_2007}, respectively.

The boundary condition at $z=0$ is inelastically reflecting so that at every bounce the particle loses some  part of its wall-normal velocity
\begin{equation}
\label{boundary_condition}
v\to -\beta v,
\end{equation}
where $0<\beta< 1$.
In general, $\beta$ can be viewed as an effective restitution coefficient related to energy dissipation due to collision-induced plastic deformations and/or particle-wall hydrodynamic interaction \cite{Legendre_2006}.
The realistic models for these dissipation processes should involve $\beta$ which is a function of the impact velocity.
Besides, the temporal or spatial irregularities at the surface of the wall may give rise to randomness of this parameter.
In this study we restrict ourselves to an idealized model with the constant restitution coefficient $\beta$.

\section{INELASTIC COLLAPSE}

\begin{figure}
\center{\includegraphics[scale=0.32]{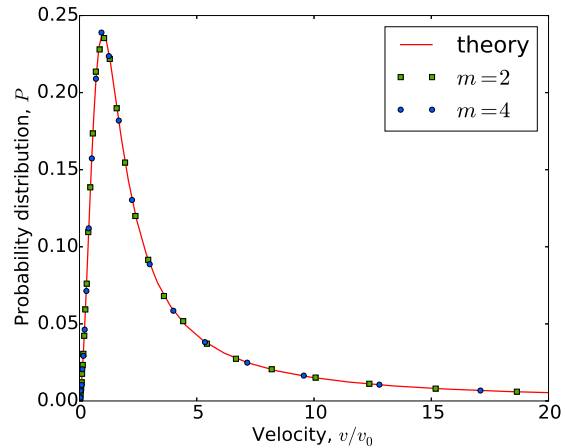}}
  \caption{Numerical results for the probability distribution of the first return velocity for different profiles (\ref{temperature_profile}) of noise intensity: $m=2$ (squares) and $m=4$ (circles). Solid line represents the theoretical prediction (\ref{first_return}). }
  \label{pic:first return}
  \end{figure}

In \cite{McKean} and \cite{Cornell} the typical trajectory of randomly accelerated particle moving near an inelastic wall was studied for the case of homogeneous forcing ($m=0$).
In particular, considering the particle leaving the wall at $z=0$ with the initial velocity $v_0>0$,  Cornell et. al. derived  analytically the statistical distribution of the return velocity $v$ right after the $n$th boundary collision \cite{Cornell}.
The corresponding probability density function of  $u=\ln v/v_0$ in the large $n$ limit is as follows:
\begin{equation}
\label{Cornell}
Q_n(u)\sim n^{-1/2}\exp\left[-\frac{9}{8n\pi^2}\left(u-n\ln\frac{\beta}{\beta_c}\right)\right]
\end{equation}
where $\beta_c=e^{-\pi/\sqrt3}\approx 0.163$.
The distribution (\ref{Cornell}) is sharply peaked around the value $n\ln(\beta/\beta_c)$ giving the typical collision velocity $v_n\sim v_0(\beta/\beta_c)^{n}$, which exponentially increases with the number of collisions at $\beta>\beta_c$, and exponentially decreases otherwise.
Then for $\beta<\beta_c$  particle trajectories undergo inelastic collapse, touching the origin of the
phase space after an infinite number of collisions with probability $1$.

Let us demonstrate that Eq. (\ref{Cornell}) remains valid in the case of inhomogeneous forcing.
The key ingredient in the original derivation of this formula by  Cornell et. al. \cite{Cornell} is the velocity distribution $P(v|v_0)$ on the first return to the origin for a particle released from $z=0$ with velocity $v_0>0$.
In general, this quantity can be expressed as $P(v|v_0)=v\rho(0,-v)$ where the probability density function  $\rho(z,v)$ corresponding to the stochastic equations (\ref{equation of motion}) is the solution of the boundary value problem:
\begin{eqnarray}
\label{FPE222}
-v\partial_z\rho+\kappa(z)\partial_v^2\rho=0,\\
\label{source}
\rho(0,v)=\frac{1}{v_0}\delta(v-v_0),\ \ \ \text{for}\ \ \ v>0.
\end{eqnarray}
Solving Eqs. (\ref{FPE222}) and (\ref{source}) for $\kappa=1$  one obtains the probability distribution for the velocity of the first return:
\begin{equation}
\label{first_return}
P(v|v_0)=\frac{3}{2\pi}\frac{v_0^{1/2}v^{3/2}}{v^3+v_0^3},
\end{equation}
which after some algebra leads to Eq. (\ref{Cornell}) (see \cite{Cornell} for the details).
In the case of inhomogeneous temperature,   the new variable $\chi=\int_{0}^z\kappa(z')dz'$, turns  (\ref{FPE222}) into $-v\partial_\chi\rho+\partial_v^2\rho=0$, while Eq. (\ref{source}) remains the same. In other words, the problem is reduced to that  with $\kappa=1$ provided $\kappa(z)$ is integrable at $z\to 0$.
Therefore, the probability distributions (\ref{Cornell}) and (\ref{first_return})  do not depend on the specific form of $\kappa(z)$.

Universality of the distribution (\ref{first_return}) for the different profiles of noise intensity is confirmed by the direct numerical simulations of the stochastic equations (\ref{equation of motion}), see Fig. \ref{pic:first return}.

Based on Eq. (\ref{Cornell}), we conclude that for a wide class of models the particle comes to the wall with zero velocity after infinite number of collisions as long as $\beta<\beta_c$.
The dynamics of the trajectory collapse, however, is non-universal.
Let us estimate how long does the collapse take for  different power-law profiles (\ref{temperature_profile}).
%For the particle starting motion towards the wall from the distance  $z$ the typical initial velocity is  $\sim  z^{(m+1)/3}$.
%The fly to the boundary is ballistic in the main approximation due to the rapid decrease in noise intensity near the wall,
%Then the typical particle velocity at the moment of collision also behaves as $z^{(m+1)/3}$.
Since the typical velocity of particle at $n$th collision is $v_n\sim v_0(\beta/\beta_c)^{n}$, then the particle get this velocity at $ z_n\sim  v_n^{3/(m+1)}$.
This estimate is found  by comparison of two terms in the rhs of the Fokker-Planck equation (\ref{FPE222}).
Therefore, time intervals between collisions can be estimated as $T_n\sim z_n/v_n$.
Next, we write $d\tilde z/dt\sim (z_{n+1}-z_{n})/T_n \sim \tilde z^{(m+1)/3}\ln(\beta/\beta_c)$, where $\tilde z(t)$ is the characteristic particle coordinate.
For $m\ne 2$ one obtains
\begin{equation}
\tilde z^{\frac{2-m}{3}}-\tilde z_0^{\frac{2-m}{3}}\sim \frac{2-m}{m+1}\ln\frac{\beta}{\beta_c}t
\end{equation}
Thus, the collapse takes infinitely long at $m>2$, and finite time at $m<2$.
Note also that in the quadratic case $m=2$ our naive estimate gives $\ln\tilde z(t)/ \tilde z_0\sim\ln(\beta/\beta_c)t$, i.e. the typical distance from the wall exhibits exponential behaviour; this result is in agreement with an exact analysis of quadratic model published before \cite{Belan_2016}.

\section{NEAR-WALL LOCALIZATION}

What can we learn about the real-space particle distribution from the collapse transition in the probability density (\ref{Cornell}) of reflected velocities?
For $m=0$, Cornell et. al.  \cite{Cornell} have argued that the particle remains at rest on the wall once its trajectory in phase space touches the origin.
% and, thus, the inelastic collapse lead to particle localization at the wall.
This statement has been questioned by many authors on the basis of simulations and theoretical analysis \cite{Florencio_2000, Anton_2002, Burkhardt_2004}.
%In particular, Burkhardt and Kotsev \cite{Burkhardt_2004} have constructed the equilibrium distribution function for a particle between two inelastic walls which remains extended (i.e. non-collapsed) for any $0<\beta\le 1$.
It has been realized that at $\beta<\beta_c$ the problem is indeterminate unless we specify
the boundary condition at the origin of phase space \cite{Anton_2002,Hwang_2015}.
If the collapsed trajectories are set to terminate (trapping condition), then the origin plays a role of an absorbing point and particle localization at $z=0$ occurs indeed.
However, if these trajectories are not set to terminate (non-trapping condition), after collapse the particle is sent back into the domain $z>0$ with unit probability.
Thus, for $m=0$, depending on the boundary condition at the origin one may obtain either collapsing or non-collapsing solution at $\beta<\beta_c$.
%Accordingly to the conventional viewpoint, once particle stop its motion at the wall it does not remain there for later times.

An open problem in the model $m=0$ (which we solve below) is whether the non-collapsing solution is localized in a vicinity of the wall or the particle eventually escapes to infinity. 
In general, it is clear that the possibility of localization for the non-trapping boundary condition must depend on the profile of noise intensity --- if it turns into zero at $z\to0$ fast enough, one may expect the particle to stay near the wall.
To address localization for arbitrary  profile of noise intensity
% assuming the case of  non-collapsing solution for which the total probability in the region $z>0$ is conserved.
we consider the equilibrium probability density function $\rho(z,v)$  which is the solution of the stationary Fokker-Planck equation
 \begin{equation}
 \label{FPE1}
 -v\partial_z \rho
 +\kappa(z) \partial_v^2 \rho=0,
 \end{equation}
 supplemented by the inelastic boundary condition
 \begin{equation}
\label{boundary cond}
 \rho(0,v)=\beta^{-2} \rho(0,-v/\beta) \quad \mathrm{for}\ v>0,
 \end{equation}
which ensures that the outcoming flux of probability
$\rho vdv$ for positive $v$  coincides with the incoming
flux for $-v/\beta$.
Near-wall localization means that the equilibrium real-space probability distribution $n(z)=\int_{-\infty}^{+\infty} \rho(z,v) dv$ is normalizable at $z\to\infty$.

In \cite{Belan_2015} the boundary value problem (\ref{FPE1}-\ref{boundary cond}) has been solved for $\kappa(z)=z^4$.
Here we extend that approach to construct $\rho$ for arbitrary temperature profile $\kappa(z)$ which is only assumed to be integrable at $z\to0$.
Let us introduce a self-similar substitution
 \begin{eqnarray}
 \rho= \chi(z)^{-a} h(\zeta),
 \label{defzeta}
 \end{eqnarray}
where
\begin{equation}
\zeta =\frac{v^3}{9\chi(z)},
 \quad
 \chi(z)=\int\limits_{0}^{z}\kappa(z')dz',
\end{equation}
and $a$ is some scaling index.
Inserting  (\ref{defzeta}) into (\ref{FPE1}), it is easy to see that unknown function $h$ satisfies the confluent hypergeometric equation
 \begin{equation}
 \label{self similar}
 \zeta\partial_\zeta^2 h
 +\left({2}/{3}+\zeta\right) \partial_\zeta h +ah=0,
 \end{equation}
whose solutions are the Kummer function $M(a,2/3,-\zeta)$ and the Tricomi function $U(a,2/3,-\zeta)$.
Since the function $M(a,2/3,-\zeta)$ diverges exponentially at large negative $\zeta$, one should choose
$h=U(a,2/3,-\zeta)$ at $\zeta<0$.
At positive $\zeta$ the function $h$ is a linear combination of $M(a,2/3,-\zeta)$ and $U(a,2/3,-\zeta)$. Equating the values of the function $h$ and of its derivative over $v$ at $\zeta=+0$ and $\zeta=-0$, one obtains
\begin{eqnarray}
\nonumber
h(\zeta>0)&=&\frac{2}{\sqrt3}\mathrm{Im}\, U(a,\frac23,-\zeta)
 +\frac{\Gamma(\frac13)}{\Gamma(a+\frac13)}
 M(a,\frac23,-\zeta),\\ % \ \zeta>0,\\
  \label{equilibrium_distribution}
 h(\zeta<0)&=&U(a,\frac23,-\zeta). 
\end{eqnarray} %, \ \zeta<0.

Next, exploiting the asymptotic behavior of $M$ and $U$ at large values of $|\zeta|$, we find
 \begin{eqnarray}
 h(\zeta \to +\infty)&\approx& \frac{2}{\sqrt3}\left\{\sin(\pi a)+
 \sin\left[\left(\frac{2}{3}-a\right)\pi\right]\right\}
 \zeta^{-a},
  \nonumber \\
  h(\zeta\to-\infty)&\approx& |\zeta|^{-a},
 \label{asymp}
 \end{eqnarray}
Since $\zeta\to\infty$ at $z\to0$, one obtains from (\ref{asymp}) the following identity for the velocity distribution at the surface of the wall
 \begin{equation}
 \rho(0,v) =\frac{2}{\sqrt3}\left\{\sin(\pi a)+
 \sin\left[\left(\frac{2}{3}-a\right)\pi\right]\right\}
 \rho(0,-v),
 \label{asymp1}
 \end{equation}
in which $v>0$.
Comparing this relation with the boundary condition (\ref{boundary cond}) we obtain the equation
 \begin{equation}
 \frac{\sin(\pi a)-\sin[\pi(a-2/3)]}{\sin(2\pi/3)}
 =\beta^{3(a-2/3)},
 \label{ccc2}
 \end{equation}
that determines the scaling index $a$ as a function of elasticity  coefficient $\beta$.

\begin{figure}
\center{\includegraphics[scale=0.32,origin=c]{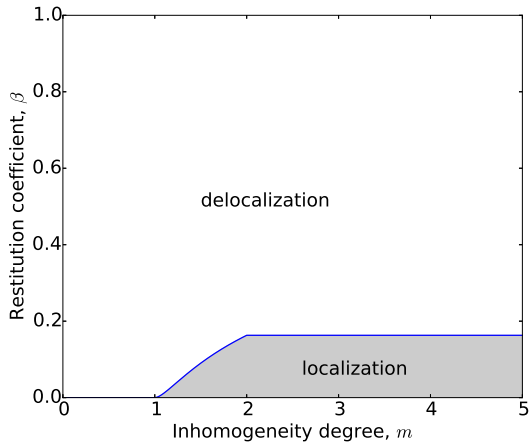}}
  \caption{Phase diagram of localization-delocalization transition for a randomly accelerated particle.}
  \label{pic:phase diagram}
  \end{figure}

The equation (\ref{ccc2}) has a solution $a=2/3$ at any $\beta$, that gives the probability density function
having symmetric tails $\rho(z,v)\propto v^{-2}$ for $v\to\pm\infty$.
However, this stationary distribution requires a particle source at the wall since it has a positive probability flux $\int dv\ v \rho$.
Indeed, multiplying the equation (\ref{self similar}) by $\zeta^{-1/3}$, integrating over $\zeta$ and assuming that $h$ tends to zero fast enough at $\zeta\to\pm\infty$, we obtain $(a-2/3)\int dv\ v \rho=0$.
Thus, the choice $a=2/3$ corresponds to nonzero flux.
The zero-flux solution of Eq. (\ref{ccc2}) starts from $a=5/6$ at $\beta=0$, then $a$ decreases as $\beta$ grows.
At $\beta\to \beta_c=e^{-\pi/\sqrt3}$ one obtains $a=2/3$.
When $\beta$ grows further, exceeding $\beta_c$, that branch gives $a<2/3$.
In this case the tails of $\rho$ at $v\to\pm\infty$ decay slower than $v^{-2}$ an condition of zero flux $\int dv\, v \rho=0$ cannot be satisfied.
We thus conclude that the equilibrium probability distribution exists only for the values of $\beta$ below the threshold of the inelastic collapse. % when $\beta<\beta_c=e^{-\pi/\sqrt3}$.

The stationary probability density for the position of the particle is $ n(z)=\int_{-\infty}^{+\infty} \rho(z,v)dv\propto \chi(z)^{-a+1/3}$,
 where the scaling index $a(\beta)$ is determined by Eq. (\ref{ccc2}) and varies from $5/6$  to $2/3$ for $0\le \beta\le\beta_c$.
For the power-law temperature profile (\ref{temperature_profile}) we obtain
\begin{equation}
 \label{concentration1}
 n(z)\propto \frac{1}{z^{(m+1)(a-1/3)}} .
 \end{equation}
The particle is localized if this distribution is normalisable at infinity.
Figure 2 represents the phase diagram of the localization-delocalization transition in $\beta-m$ plane.
Since $2/3< a\le5/6$,  distribution (\ref{concentration1}) is integrable at $z\to\infty$ for any $m\ge 2$.
Therefore, for $m\ge 2$ the particle is localized near the wall at $\beta<\beta_c$, and delocalized otherwise.
Next, for $1<m<2$ the distribution (\ref{concentration1}) is localized provided $\beta<\tilde\beta_c(m)$, where the critical value $\tilde\beta_c$ is determined by the condition   $(m+1)(a(\tilde\beta_c)-1/3)=1$.
Finally, if $0\le m\le 1$, the spatial density (\ref{concentration1})  decays at $z\to\infty$ too slowly and the particle is delocalized even in the limit of totally inelastic collisions, $\beta\to 0$ ($a=5/6$).

Thus, the region of localization coincides with the region of inelastic collapse only if the noise inhomogeneity is strong enough, namely, for $m\ge 2$.
At $0\le m<2$ one may obtain a situation when trajectories undergo inelastic collapse, but the particle is nevertheless delocalized.
It is important to remind here that for $0\le m<2$ the collapse takes a finite time.
It was proved previously for the particular case of homogeneous noise $m=0$ that after collapse the particle leaves the wall with unit probability provided the boundary condition in the origin of phase space is non-trapping (see \cite{Hwang_2015} and relevant references therein).
The existence of extended (non-collapsed) equilibrium probability distribution at $\beta<\beta_c$ indicates that escape from the wall is possible also for $0< m<2$ despite the fact that the intensity of the random force vanishes at $z=0$. Then, the delocalization  may occur even at $\beta<\beta_c$ due to long-distance excursions between collapse events.

\section{VISCOUS DUMPING}

Finally, let us examine how the presence of viscous damping change the localization properties of a randomly accelerated particle. 
Now the equations of motion is
\begin{equation}
\label{equation of motion viscous}
\frac{dz}{dt}=v,\ \ \ \ \frac{dv}{dt}=-\frac{v}{\tau}+\xi,
\end{equation}
where $\tau$ is the velocity relaxation time.
The  Fokker-Planck equation is as follows
 \begin{equation}
 \label{FPE3}
 -v\partial_z \rho + \frac{1}{\tau}\partial_v(v\rho) +\kappa(z) \partial_v^2 \rho=0,
 \end{equation}
As previously,  we should impose the boundary condition (\ref{boundary cond}) to take into account the inelastic collisions at the wall.
%Physically, the viscosity provides another mechanism of energy dissipation in addition to inelastic collisions.
%
%It is instructive to note that Eq. (\ref{FPE3}) is equivalent to the Fokker-Planck equation for a Brownian particle in the thermal bath with inhomogeneous temperature.

Particle is localized if there exist an equilibrium probability distribution $\rho(z,v)$, and the corresponding real-space  probability density $n(z)$ is normalisable at $z\to\infty$. 
%We have not been able to construct an exact solution of the boundary-value problem  consisting of Eqs. (\ref{FPE3}) and (\ref{boundary cond}).
%However, to establish whether the particle is localized or not, it is enough to find the large-$z$ tail of the equilibrium probability density $n(z)$ of particle coordinate.
While we have not been able to construct an exact solution of the boundary-value problem  consisting of Eqs. (\ref{FPE3}) and (\ref{boundary cond}, the large-$z$ tail of $n(z)$ can be easily found.

The equilibrium state (if any) is characterized by a balance between the velocity fluctuations imposed on the particle by the random force and the dissipation of those fluctuations due to viscous friction and inelastic collisions.
To measure the %relative contribution of viscous dumping in
role of temperature inhomogeneity in 
formation of the equilibrium velocity distribution at given distance from the wall, we introduce the following dimensionless parameter 
\begin{equation}
\label{inertia_degree}
I(z)=\frac{\tau}{\tilde\tau(z)},
\end{equation}
where $\tilde\tau(z)$ represents the local time scale given by the time it takes the particle to
experience inhomogeneity of environment.
We can estimate $\tilde \tau(z)$ as the characteristic scale of inhomogeneity, which is just $z$ in the case of scale-free profiles (\ref{temperature_profile}), divided by the typical velocity $\tilde v(z)$ of the particle, i.e. $\tilde\tau(z) =z/\tilde v(z)$.
Equivalently,  parameter $I$ can be  defined as the ratio of the mean free path $l(z)\sim \tilde v(z)\tau$ and the distance to the wall $z$.
%Let $\tilde v(z)$ be the typical velocity of the particle at given distance from the wall.  
%Then we can define 
%Equivalently, this parameter can be define as the local estimate for the mean path $l\sim \tilde v(z)\tau$ divided by the distance to the wall $z$.
In those regions where $I(z)\ll 1$, the fluctuation dissipation mechanism operates locally: if the velocity distribution at a given point is perturbed, it will relax to equilibrium via viscous dumping essentially experiencing the same temperature. 
In contrast, when $I(z)\gg 1$, the statistics of particle velocity is far from local equilibrium with noise
% with noise because particle feels the temperature inhomogeneity for the time which is much smaller than viscous relaxation time $\tau$.
%In the later case 
since the particle can reach the wall for the time which is much smaller than velocity relaxation time $\tau$. 
In this case the velocity fluctuations are balanced mainly by dissipative wall collisions.

\begin{figure}
\center{\includegraphics[scale=0.32,origin=c]{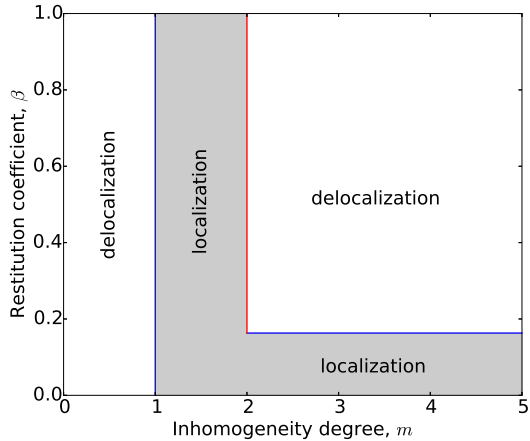}}
  \caption{Phase diagram of localization-delocalization transition for a randomly accelerated particle in the presence of viscous dumping. Along the red line the localization properties depend also on the parameter $I$, see \cite{Belan_2016}. }
  \label{pic:phase diagram dumping}
  \end{figure}
%the particle gain some energy from the noise,

Let us assume that condition $I(z)\ll 1$ is valid.
Then the typical velocity $\tilde v(z)$ is determined by the interplay of the second and the third terms in the rhs of Eq. (\ref{FPE3}): $\tilde v(z)\sim \sqrt{\kappa(z)/\tau}$.
%Then the typical velocity can be estimated simply by comparing the second and the third terms in the rhs of Eq. (\ref{FPE3}): $\tilde v\sim \sqrt{\gamma^{-1}\kappa}$, and, therefore, $\tilde\tau\sim$.
Using this estimate we obtain the following $z$-dependence for the parameter (\ref{inertia_degree})
\begin{equation}
I(z)\propto z^{m/2-1}.
\end{equation}
%Therefore, the assumption of local equilibrium is justified when $z^{m/2-1}\ll \tau^{-3/2}$.
One needs to consider separately three different cases: $0\le m<2$, $m>2$ and $m=2$.

\subsection{$0\le m<2$}

If $0\le m<2$, then $I(z)\to 0$ at $z\to\infty$. 
By other words, the typical free path of the particle at $z\to\infty$ is much smaller than distance to the wall.
%dynamics does not have enough time to experience the variations in spacial statistics of the noise. 
Therefore, 
%the particle feels only its local temperature in the main approximation: 
the approximation of local equilibrium is justified:
the statistics of particle velocity at large distance from the wall is determined by the local temperature of noise. 
%the body of velocity distribution at large distance from the wall is formed by the particles which are in statistical equilibrium with local noise.
%  starting from an arbitrary initial velocity distribution, a local quasi-equilibrium velocity distribution will be
%established after a time of the order $\tau$ for every $z$.
%the the body of the velocity distribution is formed by the particles which are in statistical equilibrium with the local temperature.
%That limit has been studied intensively in the fields of overdamped Brownian motion in non-isothermal bath [?] and turbulent transport of inertial particles in wall-bounded flows. 
Performing adiabatic elimination of velocity in Eq. (\ref{FPE3}) one obtains the following  gradient transport equation \cite{Belan_2014, Belan_2016_1}
 \begin{equation}
\partial_t n=\tau^2\partial_z^2[\kappa n],
 \end{equation}
which describes evolution of the real-space probability distribution $n(z,t)$ on timescales $t\gg\gamma^{-1}$ and at large $z$.
It is straightforward to find the equilibrium solution of this equation  for $\kappa$ given by (\ref{temperature_profile}):
\begin{equation}
\label{local equilibrium}
n(z)\propto \frac{1}{z^{m}}.
\end{equation}
If $1<m<2$, then the equilibrium solution is normalisable at infinity and, thus, the particle is localized.
For $0\le m \le 1$, the solution decays too slowly and the particle is delocalized.  
Note that localization properties are $\beta$-independent since at $z\to\infty$ the particle becomes insensitive to the boundary conditions at the wall. 
%At the same time the probability distribution at small distances from the wall depends on the restitution coefficient. 

\subsection{$m>2$}

For $m>2$, we have $I(z)\to\infty$ at $z\to \infty$.
%the local equilibrium is strongly violated at $z\to \infty$, since
The particle placed at arbitrary large distance from the wall 
% the body of the velocity distribution at large distance from the wall is formed by the particles which 
 can reach the wall in ballistic manner and for this reason the local equilibrium is strongly violated. 
It can be shown that the viscous dumping  is asymptotically negligible at $z\to \infty$: the region of phase space where the second term in the rhs of Eq. (\ref{FPE3})  is important becomes very narrow (along $v$-axis) in comparison with the body of the velocity distribution (see \cite{Belan_2015} for more details).
%typical free path is much larger than distance to the wall.
%Then viscous dumping is asymptotically negligible and the main mechanism of energy dissipation at large $z$ is inelastic collisions. 
%At $z\to \infty$ the region of phase space where the second term in the right hand side of Eq. (\ref{FPE3})  is important becomes very narrow (along $v$-axis) in comparison with the body of the velocity distribution (see \cite{Belan_2015} for more details).
Therefore, to describe approximately the probability distribution $\rho(z,v)$ at large distances from the wall we can pass to Eq. (\ref{FPE1}).
Moreover, the approximate solution can be drawn to $z\to 0$ (that corresponds to ballistic flights of the particle to the wall and back)  and, thus, it must satisfy the boundary condition (\ref{boundary cond}). 
All this means that the asymptotic behavior  of $n(z)$ at large $z$ is given by Eq. (\ref{concentration1}). 
We conclude that the presence of viscous damping does not change the phase diagram for $m>2$: the particle is localized if $\beta<\beta_c$, and is delocalized otherwise.

%To justify the zero-viscosity approximation more rigorously let us  at $z\to 0$ it is enough to notice that
%
%More formally, comparing the terms in the right hand of Eq. (\ref{FPE3}) we conclude that "viscous" term is negligible provided $z\ll$ and $|v|\gg z/\tau$.
%The interplay of the remaining terms gives the following estimate for the typical velocity: $\tilde v\sim z^{(m+1)/3}$.
%As $z\to \infty$, the region of phase space where viscosity is relevant becomes very narrow (along $v$-axis) in comparison with the body of the velocity distribution.
%Therefore, we can pass to Eq. (\ref{FPE1}) to describe the probability density at large distances from the wall. 
%Importantly, the large-$z$ solution 
%
%
%Therefore we can consider the solution  large-$z$ tail of $n$ can be constructed using Eq.   
%Moreover, the viscosity 
%the probability distribution gives an approximate solution of at $z\to 0$. 

\subsection{$m=2$}

In the marginal case $m=2$, parameter $I$ becomes $z$-independent and can be interpreted as a measure of particle inertia \cite{Belan_2016}.
If $I\to 0$, then the locally-equilibrium approach gives the profile  $n(z)\propto 1/z^2$, which is normalisable at $z\to\infty$, i.e. the particle is localized for any $\beta$.
In the opposite limit $I\to \infty$, the zero-friction approximation is justified and one obtains localization-delocalization transition at $\beta=\beta_c$ as it follows from the results of the previous section.
%In general, there is no way to determine the solution of Eq. () since all term should be taken into account.
It is possible also to describe the localization properties for any value of the parameter $I$.
For this aim we should find the sign of the Lyapunov exponent which is defined as $\lambda=\lim_{t\to\infty}t^{-1}\langle\ln z(t)/z(0)\rangle$.
The negative Lyapunov exponent corresponds to localization: the particle coordinate exponentially decreases with time.
In contrast, the positive sign of the Lyapunov exponent means delocalization because the particle coordinate  exponentially grows. 
Exact calculation of $\lambda$ allows to determine the phase curve of the localization-delocalization transition in $\beta-I$ plane (see Fig.~2 in \cite{Belan_2016}).
Note that the particle is localized for any $I$ when $\beta<\beta_c$.
Figure \ref{pic:phase diagram dumping} summarizes the results of this section.

%The particle moves via diffusion mechanism and does not sense the boundary conditions at the wall in the leading-order approximation. 

\section{CONCLUSION}

We have studied the one-dimensional dynamics of a randomly accelerated particle colliding with inelastic boundary at $z=0$.
In our model the effective temperature $\kappa$ of noise may depend on the spatial coordinate $z$; the main focus is on the power-law temperature profiles $\kappa=z^m$, where $m\ge 0$, motivated by the fluid-mechanical applications.
A surprising result of our study is universality of the collapse transition which was previously known only for the particular case $m=0$: at $\beta>\beta_c=e^{-\pi/\sqrt 3}$ the typical velocity at the moment of $n$th collision exponentially grows with $n\to \infty$, but for $\beta<\beta_c$ the collision velocity exponentially decreases.
Thus, if the velocity restitution coefficient is small enough, then the particle finds itself on the wall with zero velocity after an infinite number of inelastic collisions.  
This conclusion is valid for any $z$-dependence of $\kappa$ provided it is integrable at $z\to 0$.
For $\kappa=z^m$ with $0\le m<2$, the collapse occurs in a finite time.

Another important result of this paper is the exact equilibrium solution $\rho(z,v)$ of the Fokker-Planck equation with non-uniform temperature, which has the universal self-similar form (\ref{defzeta}).
Note, the solution exists only if $\beta<\beta_c$, i.e. the equilibrium state is possible only below the collapse threshold. 
On the other hand, the solution is extended in space indicating that
% the particle can leave the wall after the trajectory collapse. 
 in those cases when infinite number of collisions occurs for a finite time the particle can leave the wall after the trajectory collapse. 
Thus, the inelastic collapse does not necessarily localize the particle at
the wall.
Furthermore, based on the normalization properties of the equilibrium probability distribution (\ref{concentration1}) of particle coordinate, we have concluded that inelastic collapse does not guaranty even the near-wall localization, see Fig.~\ref{pic:phase diagram}.
For  $m\ge 2$, the particle indeed becomes localized in vicinity of the boundary as the restitution coefficient $\beta$ is getting smaller than $\beta_c$.
However, if $1<m<2$, then %
% in the non-trapping case 
 the particle is localized only at $\beta<\tilde\beta_c(m)$, where  $0<\tilde\beta_c<\beta_c$.
Finally, in the models with $0\le m\le 1$ one obtains delocalization for any $\beta$.
% provided the non-trapping condition at the origin of phase space.

We have also analyzed how the viscous dumping modifies the phase diagram of the localization-delocalization transition, see Fig.~\ref{pic:phase diagram dumping}.
Whether or not the particle is localized does not depends on the presence of viscous friction when $0\le m\le 1$ or $m> 2$.
For $1<m<2$, however, the friction leads to localization at any value of the restitution coefficient $\beta$. 
In marginal case $m=2$, the particle is always localized if $\beta<\beta_c$, while for $\beta>\beta_c$
the answer depends on the dimensionless parameter $I$ which measures the particle inertia.

The results summarized above raise several questions for the future studies.
Here we have focused on the non-collapsing solution implying that after collapse the particle does not remain at rest at the wall and re-enter the region $z>0$.
However, in practice, the  particles can be trapped in a surface potential well provided its impact velocity is small enough.
This case corresponds to the trapping boundary condition at the origin phase space, i.e. the particle trajectory should be set to terminate there.
It  might be interesting to calculate the rate at which particles deposit on the wall due to the inelastic collapse.
At the moment, the answer is known only for the homogeneous model, see \cite{Swift_1999, Burkhardt_2000_2}.
An even more interest question concerns the  higher-dimensional generalization of the present analysis.
Is there a collapse transition for an ensemble of inelastically colliding particles placed in turbulent flow in two or three spatial dimensions?
That issue is of great importance for a variety of applications ranging from cloud physics to planet formation.

{}

\end{document}